\documentclass[]{aa} 
\usepackage{graphicx,hyperref,color}
\usepackage{txfonts}
\usepackage{longtable,lscape}
\begin{document}
   \title{High rigidity Forbush decreases: due to CMEs or shocks?}


   \author{Arun Babu\inst{1}, H. M. Antia\inst{2}, S. R. Dugad\inst{2}, S. K. Gupta\inst{2}, Y. Hayashi\inst{3}, S. Kawakami\inst{3}, P. K. Mohanty\inst{2}, T. Nonaka\inst{3}, A. Oshima\inst{4}, P. Subramanian\inst{1} \\{\large (The GRAPES-3 Collaboration)}
           }
   \institute{Indian Institute of Science Education and Research, Sai Trinity Building, Pashan, Pune 411 021, India \and
Tata Institute of Fundamental Research, Homi Bhabha Road,
Mumbai 400 005, India \\
The GRAPES--3 Experiment, Cosmic Ray Laboratory, Raj Bhavan, Ooty 643 001, India \and
Graduate School of Science, Osaka City University, Osaka 558-8585, Japan           \and 
National Astronomical Observatory of Japan, CfCA, Tokyo 181-8588, Japan
}
\authorrunning{Arun Babu et al.}
\titlerunning{High rigidity Forbush decreases}
   \date{}

 
  \abstract
   {}
{We seek to identify the primary agents causing Forbush decreases (FDs) observed at the Earth in high rigidity cosmic rays. In particular, we ask if such FDs are caused mainly by coronal mass ejections (CMEs) from the Sun that are directed towards the Earth, or by their associated shocks.}
{We use the muon data at cutoff rigidities ranging from 14 to 24 GV from the GRAPES-3 tracking muon telescope to identify FD events. We select those FD events that have a reasonably clean 
profile, and can be reasonably well associated with an Earth-directed CME and its associated shock. We employ two models: one that considers the CME as the sole cause of the FD (the CME-only model) and one that considers the shock as the only agent causing the FD (the shock-only model). We use an extensive set of observationally determined parameters for both these models. The only free parameter in these models is the level of MHD turbulence in the sheath region, which mediates cosmic ray diffusion (into the CME, for the CME-only model and across the shock sheath, for the shock-only model).}
{We find that good fits to the GRAPES-3 multi-rigidity data using the CME-only model require turbulence levels in the CME sheath region that are only slightly higher than those estimated for the quiet solar wind. On the other hand, reasonable model fits with the shock-only model require turbulence levels in the sheath region that are an order of magnitude higher than those in the quiet solar wind.}
   {This observation naturally leads to the conclusion that the Earth-directed CMEs are the primary contributors to FDs observed in high rigidity cosmic rays.}


   \maketitle

\section{Introduction}

Forbush decreases are short-term decreases in the intensity of the galactic cosmic rays at the Earth. They are typically caused by the effects of interplanetary counterparts of coronal mass ejections (CMEs) from the Sun, and also the corotating interaction regions (CIRs) between the fast and slow solar wind streams from the Sun. We concentrate on CME driven Forbush decreases in this paper. The near-Earth manifestations of CMEs from the Sun typically have two major components: the interplanetary counterpart of the CME, commonly called an ICME, and the shock which is driven ahead of it.
ICMEs which possess certain well-defined criteria such as plasma
temperature depressions and smooth magnetic field rotations (e.g., Burlaga et al. 1981, Bothmer \& Schwenn 1998) are called magnetic clouds, while others are often classified as ``ejecta''. The relative contributions of shocks and ICMEs in causing Forbush decreases is a matter of debate. For instance, Zhang \& Burlaga (1988), Lockwood, Webber, \& Debrunner (1991) and Reames, Kahler \& Tylka (2009) argue against the contribution of magnetic clouds to Forbush decreases. On the other hand, other studies (e.g., Badruddin, Yadav, \& Yadav, 1986; Sanderson et al.,1990; Kuwabara et al 2009) concluded that magnetic clouds can make an important contribution to FDs. There have been recent conclusive associations of Forbush decreases with Earth-directed CMEs (Blanco et al 2013; Oh \& Yi 2012).  Cane (2000) introduced the concept of a ``2-step'' FD, where the first step of the decrease is due to the shock and the second one is due to the ICME. Based on an extensive study of ICME-associated Forbush decreases at cosmic ray energies between 0.5 -- 450 MeV, Richardson and Cane (2011) conclude that shock and ICME effects are equally responsible for the Forbush decrease. They also find that ICMEs that can be classified as magnetic clouds are usually involved in the largest of the Forbush decreases they studied. From now on, we will use the term ``CME'' to denote the CME near the Sun, as well as its counterpart observed at the Earth.

In this paper we have used Forbush decrease data from GRAPES-3 tracking muon telescope located at Ooty (11.4$^{\circ}$N latitude, 76.7$^{\circ}$E longitude, and 2200~m altitude) in south India. The GRAPES-3 muon telescope records the flux of muons in nine independent directions (labeled NW, N, NE, W, V, E, SW, S and SE), and the geomagnetic cutoff rigidity over this field of view varies from 12 to 42 GV. The details of this telescope are discussed in Hayashi et al. (2005), Nonaka et al. (2006) and Subramanian et al (2009). Thus, the GRAPES-3 telescope observes the cosmic ray muon flux in nine independent directions with varying cut-off rigidities simultaneously. 
The high muon counting rate measured by the GRAPES-3 telescope results in extremely small statistical errors, allowing small changes in the intensity of the cosmic ray flux to be measured with high precision. Thus a small drop ($\sim$0.2\%) in the cosmic ray flux during a Forbush decrease event can be reliably detected. This is possible even in the presence of the diurnal anisotropy of much larger magnitude ($\sim$1.0\%), through a filtering technique described in Subramanian et al (2009) (referred to hereafter as S09).

\begin{figure}
\centering 
\includegraphics[width = \columnwidth]{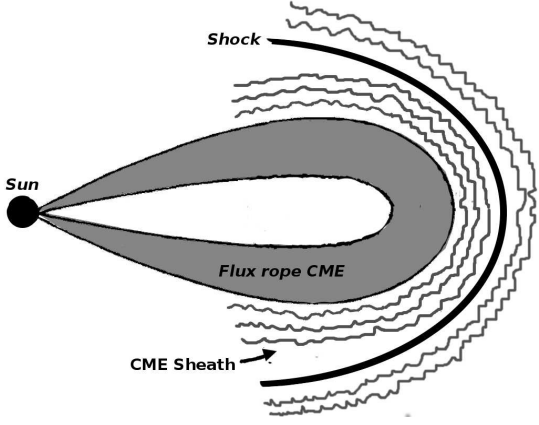}
\caption{A schematic of the CME-shock system. The CME is modeled as a flux rope structure. The undulating lines ahead of the shock denote MHD turbulence driven by the shock, while those in the CME sheath region denote turbulence in that region.}
\label{difn}
\end{figure}

A schematic of the CME, which is assumed to have a flux-rope geometry (Vourlidas et al 2012) together with the shock it drives, is shown in Figure~\ref{difn}. The shock drives turbulence ahead of it, and there is also turbulence in the CME sheath region (e.g., Manoharan et al 2000; Richardson \& Cane 2011). 

Instead of treating the entire system shown in Figure~\ref{difn}, which would be rather involved, we consider two separate models. The first, that we call the ``CME-only'' model, is one where the Forbush decrease is assumed to be exclusively due to the CME, which is progressively populated by high energy cosmic rays as it propagates from the Sun to the Earth. The preliminary idea behind this model was first sketched by Cane, Richardson \& Wibberenz (1995) and developed in detail in S09. The work described here addresses multi-rigidity data, which is a major improvement over S09. We will describe several other salient improvements in the CME-only model in subsequent sections. The second model, which we call the ``shock-only'' model, is one  where the Forbush decrease is assumed arise only due to a propagating diffusive barrier, which is the shock driven by the CME (e.g., Wibberenz et al 1998). The diffusive barrier would act as a shield for the galactic cosmic ray flux, resulting in a lower cosmic ray density behind it. In treating these two models separately, we aim to identify which is the dominant contributor to the observed Forbush decrease; the CME, or the shock.

We identify Forbush decreases in the GRAPES-3 data that are associated with both near-Earth magnetic clouds and possibly shocks driven by them. We describe our event shortlisting criteria in the next section. We then test the extent to which each of the models (the CME-only and the shock-only model) satisfies the multi-rigidity Forbush decrease data from the GRAPES-3 muon telescope. In the subsequent analysis the use of cutoff rather than median rigidity is preferred for the following reason. The cutoff rigidity in a given direction represents the threshold rigidity of incoming cosmic rays and the magnitude of Forbush decrease is a sensitive function of it. On the other hand the median rigidity is comparatively insensitive to the magnitude of the Forbush decrease.

\section{Event shortlisting criteria} \label{ESC}

\begin{table*}
\caption {Selected Events }\label{SL3}
\centering
\begin{tabular}{lcccccc}
\hline \hline
Event & Magnitude(Ver)  ($\%$) & FD onset (UT) &  FD min (UT)  & MC start (UT) & MC stop (UT)& CME launch (UT)\\ \hline
11/04/01 & 2.86 & 11/04/01, 12:00 & 12/04/01, 18:00 & 11/04/01, 23:00 & 12/04/01, 18:00 & 10/04/01, 05:30\\ 
17/08/01 & 1.03 & 16/08/01, 22:34 & 18/08/01, 05:00 & 18/08/01, 00:00 & 18/08/01, 21:30 & 15/08/0 1, 23:54\\
29/09/01 & 2.44 & 29/09/01, 12:43 & 31/09/01, 01:00 & 29/09/01, 16:30 & 30/09/01, 11:30 & {\ldots}not clear{\ldots}  \\
24/11/01 & 1.56 & 24/11/01, 03:21 & 25/11/01, 15:00 & 24/11/01, 17:00 & 25/11/01, 13:30 & 22/11/01, 22:48\\
23/05/02 & 0.93 & 23/05/02, 02:10 & 23/05/02, 23:00 & 23/05/02, 21:30 & 25/05/02, 18:00 & {\ldots}not clear{\ldots}\\
 7/09/02 & 0.97 &  7/09/02, 14:52 &  8/09/02, 13:00 &  7/09/02, 17:00 &  8/09/02, 16:30 & 05/09/01, 16:54\\
30/09/02 & 0.97 & 30/09/02, 11:30 & 31/09/02, 08:00 & 30/09/02, 22:00 &  1/09/02, 16:30 & {\ldots}not clear{\ldots} \\
20/11/03 & 1.16 & 20/11/03, 10:48 & 24/11/03, 04:00 & 21/11/03, 06:10 & 22/11/03, 06:50 & 18/11/03, 08:50\\
26/07/04 & 2.13 & 26/07/04, 15:36 & 27/07/04, 11:00 & 27/07/04, 02:00 & 28/07/04, 00:00 & 25/07/04, 14:54 \\ \hline
\end{tabular}
\tablefoot{Events that can clearly be associated with near-Sun CMEs and agree with the composite velocity profile (\S~2.2) have a CME launch time (near the Sun) mentioned against them. These events form our final shortlist (shortlist 3).\\
{Magnitude(Ver) : Magnitude of Forbush decrease in vertical direction  }\\
{FD onset : Time of FD onset in UT}\\
{FD minimum : Time of FD minimum in UT}\\
{MC start : Magnetic cloud start time in UT  }\\
{MCstop : Magnetic cloud stop time in UT}\\
{CME near Sun : Time at which CME was first observed in the LASCO FOV}
}
\end{table*}

\subsection{First and second shortlists: characteristics of the Forbush decrease}
We have examined all Forbush decrease events observed by the GRAPES-3 muon telescope from 2001 to 2004. We then shortlisted events that have magnitudes $> 0.25$~\% and possess a `relatively clean' profile comprising a sudden decrease followed by a gradual exponential recovery. While the figure of 0.25 \% might seem rather small by neutron monitor standards, we emphasize that the largest events observed with GRAPES-3 have magnitudes of $\sim 1$~\%. This list, which contains 72 events, is called shortlist 1. We next correlate the events in shortlist 1 with lists of magnetic clouds near the Earth observed by the WIND and ACE spacecrafts (Huttunen et al 2005; Lynch et al. 2003; A. Lara, private communication) and select only those that can be reasonably well connected with a near-Earth magnetic cloud and this set is labeled shortlist 2 (Table~\ref{SL3}). The decrease minimum for most of the FD events in shortlist 2 lie between the start and the end of the near-Earth magnetic cloud.

\subsection{Third shortlist: CME velocity profile} \label{VP}
Since we are looking for Forbush decrease events that are associated with shocks as well as CMEs, we examine the CME catalog (\href{http://cdaw.gsfc.nasa.gov/CME_list/}{$http://cdaw.gsfc.nasa.gov/CME \_ list /$}) for a near-Sun CME that can reasonably correspond to the near-Earth magnetic cloud that we associated with the Forbush decrease in forming shortlist 2. 
In S09 it was assumed that the CME propagates with a constant speed from the Sun to the Earth. In this paper, we adopt a more realistic, two-step velocity profile that is described below. 

The data from the LASCO coronograph aboard the SOHO spacecraft (\href{http://cdaw.gsfc.nasa.gov/CME_list/}{$http://cdaw.gsfc.nasa.gov/CME \_ list /$}) provide details about CME propagation up to a distance of $\approx$   30 ${\rm R_{\odot}}$ from the Sun. We fit the following velocity profile to the LASCO data points:
\begin{equation}
\rm V_{1} = v_{i} + a_{i} t \, , \,\,\,\,\,\,\,\,\,{\rm for}\,\,\,  R(t) \leq R_m  
\label{VP1}
\end{equation}
where ${\rm v_i}$ and ${\rm a_i}$ are the initial velocity and acceleration of CME respectively, and $R_{m}$ is the heliocentric distance at which the CME is last observed in the LASCO field of view. 
For distances $> {\rm R_{m}}$, we assume that the CME dynamics are governed exclusively by the aerodynamic drag it experiences due to momentum coupling with the ambient solar wind. For heliocentric distances $> {\rm R_{m}}$, we therefore use the following widely used 1D differential equation to determine the CME velocity profile: (e.g., Borgazzi et al 2009)
\begin{equation}
\rm m_{CME} V_{2} \frac {\partial V_{2}}{\partial R}  = \frac{1}{2} C_D \rho_{sw} A_{CME} (V_{2} - V_{sw})^2 \, , \,\,\,\,\,R(t) > R_{m}
\label{VP2}
\end{equation}

\begin{figure}
\centering 
\includegraphics[scale = .85]{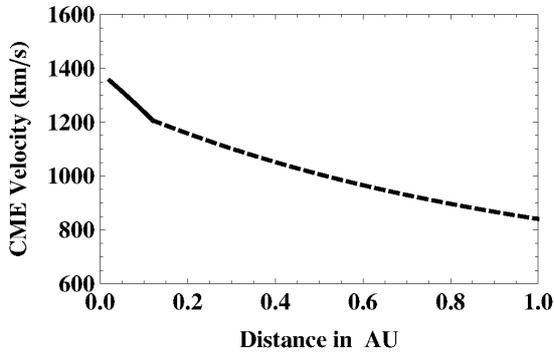}
\caption{A plot of the velocity profile for the CME corresponding to the 24 November 2001 FD event. The CME was first observed in the LASCO FOV on 22 November 2001. The solid line shows the first stage governed by LASCO observations (Eq~\ref{VP1}), where the CME is assumed to have a constant deceleration. The dashed line shows the second stage (Eq~\ref{VP2}), where the the CME is assumed to experience an aerodynamic drag characterized by a constant ${\rm C_{D}}$.}
\label{velpro}
\end{figure}

where ${\rm m_{CME}}$ is the CME mass, ${\rm C_D}$ is the dimensionless drag coefficient, ${\rm \rho_{sw}}$ is the solar wind density, ${\rm A_{CME} = \pi R_{CME}^2}$ is the cross-sectional area of the CME and ${\rm V_{sw}}$ is the solar wind speed. The boundary condition used is ${\rm V_{2} = v_{m}}$ at ${\rm R(t) = R_{m}}$. The CME mass ${\rm m_{CME}}$ is assumed to be ${\rm 10^{15}}$ g and the solar wind speed ${\rm V_{sw}}$ is taken to be equal to 450 km/s. The solar wind density ${\rm \rho_{sw}}$ is given by the model of Leblanc et al (1998). The composite velocity profile for the CME is defined by
\begin{equation}
\nonumber
\rm V_{CME} = \cases{V_{1}\, , & $R(t) \leq R_{m}$ \cr
\noalign{\medskip}
V_{2}\, , & $R(t) > R_{m}$\cr}
\label{VP3}
\end{equation}
The total travel time for the CME is ${\rm \int_{R_{i}}^{R_{f}} dR/V_{CME}}$, where ${\rm R_{i}}$ is the heliocentric radius at which the CME is first detected and ${\rm R_{f}}$ is equal to 1 AU. We have used a constant drag coefficient ${\rm C_D}$ and adjusted its value so that the total travel time thus calculated matches the time elapsed between the first detection of the CME in the LASCO FOV and its detection as a magnetic cloud near the Earth. We have retained only those events for which it is possible to find a constant ${\rm C_{D}}$ and this criterion is satisfied.
We have also eliminated magnetic clouds that could be associated with multiple CMEs.  This defines our final shortlist, which we call shortlist 3.The events that have an entry in the last column labeled `CME launch' in Table~\ref{SL3} comprise shortlist 3.  We note that we have adjusted the parameter ${\rm C_{D}}$ so that the final CME speed obtained from Eq~(\ref{VP2}) is close to the observed magnetic cloud speed near the Earth. It is usually not possible to find a ${\rm C_{D}}$ that will yield an exact match for the velocities as well as the total travel times (e.g., Lara et al 2011). Figure~\ref{velpro} shows an example of the composite velocity profile (given by Eqs~\ref{VP1} and \ref{VP2}) for one such representative CME: the one that was first observed in the LASCO FOV on 22 November 2001, and resulted in a FD on 24 November 2001.


\section{Models for Forbush decreases} \label{Mod}
We apply two different models to the FD events in Table~\ref{SL3}; the first one is the CME-only model, which assumes that the Forbush decrease owes its origin only to the CME. The second one is the shock-only model, which assumes that the Forbush decrease is exclusively due to the shock, which is approximated as a diffusive barrier. Although both the shock and the CME are expected to contribute to the Forbush decrease, our treatment seeks to determine which one of them is the dominant contributor at rigidities ranging from 14 to 24 GV. Before describing the models, we discuss the perpendicular diffusion coefficient. 



\subsection{Perpendicular diffusion coefficient}
We use an isotropic perpendicular diffusion coefficient $  ( \rm D_{\perp}) $ to characterize the penetration of cosmic rays across large scale, ordered  magnetic fields. We envisage a CME, which has a flux rope structure, propagates outwards from the Sun, driving a shock ahead of it (see, e.g., Vourlidas et al 2012). The flux rope CME-shock geometry is illustrated in Figure~\ref{difn}. The CME sheath region between the CME and the shock is known to be turbulent (e.g., Manoharan et al 2000) and it is well accepted that it has a significant role to play in determining Forbush decreases (Badruddin  2002; Yu et al 2010). The isotropic perpendicular diffusion coefficient ${\rm D_{\perp}}$ characterizes the penetration of cosmic rays through the ordered, compressed large-scale magnetic field near the shock, as well as across the ordered magnetic field of the flux rope CME. In diffusing across the shock, the cosmic rays are affected by the turbulence ahead of the shock, and in diffusing across the magnetic fields bounding the flux rope CME, the cosmic rays are affected by the turbulence in the CME sheath region.

 The subject of charged particle diffusion across field lines in the presence of turbulence is a subject of considerable research. Analytical treatments include the so-called ``classical'' scattering theory (e.g., Giacalone \& Jokipii (1999) and references therein), and the nonlinear guiding center theory (Matthaeus et al 2003; Shalchi 2010) for perpendicular diffusion. Numerical treatments include Giacalone \& Jokipii (1999), Casse et al (2002), Candia \& Roulet (2004) and Tautz \& Shalchi (2011). For our purposes, we seek a concrete, usable prescription for the  $  \rm D_{\perp} $  that can incorporate as many observationally determined quantities as possible. We find that the analytical fits to extensive numerical simulations provided by Candia \& Roulet (2004) best suit our requirements. We note, in particular, that their results not only reproduce the standard results of Giacalone \& Jokipii (1999) and Casse et al (2002) but also extend the regime of validity to include strong turbulence and high rigidities. Our approach is similar to that of Effenberger et al (2012), who adopt empirical expressions for the perpendicular diffusion coefficients. It may be mentioned, however, that they allow for the possibility of anisotropic perpendicular diffusion. We treat the case of radial diffusion across the largely azimuthal magnetic fields bounding a flux rope CME structure, and we therefore need only one (isotropic) perpendicular diffusion coefficient  $ ( \rm D_{\perp}) $.

In the formulation of Candia \& Roulet (2004), the perpendicular diffusion coefficient ${\rm D_{\perp}}$ is a function of   the quantity ${\rm \rho}$ (which is closely related to the rigidity and indicates how tightly the proton is bound to the magnetic field) and the level of turbulence ${\rm \sigma^{2}}$. Our characterization of ${\rm D_{\perp} (\rho, \sigma^{2})}$ follows that of S09, who adopt the analytical fits  to Monte Carlo simulations of particle transport in turbulent magnetic fields  given by Candia \& Roulet (2004). The quantity $\rho$ is related to the rigidity $Rg$ by
\begin{equation}
\rm \rho = \frac{R_L}{L_{\rm max}} = \frac {Rg}{B_0 L_{\rm max}}
\label{E11}
\end{equation}

where $R_{L}$ is the Larmor radius, ${\rm B_0}$ is the strength of the relevant large-scale magnetic field. For the CME-only model, $B_{0}$ refers to the large-scale magnetic field bounding the CME, and for the shock-only model, it refers to the enhanced large-scale magnetic field at the shock.
In writing second step in Eq~(\ref{E11}), we have related the Larmor radius to the rigidity Rg by

\begin{equation}
\rm R_L(t) = \frac{Rg}{B_0}\, .
\label{rl}
\end{equation}

For the CME-only model, we adopt ${\rm L_{\rm max}=2\,R(T)}$, where ${\rm R(T)}$ is the radius of the near-Earth magnetic cloud. This is in contrast with S09, where ${\rm L_{\rm max}}$ was taken to be ${\rm 10^6}$ km, which is the approximate value for the outer scale of the turbulent cascade in the solar wind. For the shock-only model, on the other hand, we assume that ${\rm L_{\rm max}}$ is equal to 1 AU.

 The turbulence level ${\rm \sigma^2}$ is defined (as in Candia \& Roulet [2004] and S09) to be 
\begin{equation}
 \rm \sigma^2 \equiv \frac{\langle {B_r}^2 \rangle}{{B_0}^2}
\label{E12}
\end{equation}

where ${\rm B_r}$ is the fluctuating part of the turbulent magnetic field and the angular braces denote an ensemble average.

 For the sake of completeness, we reproduce the full expression for the isotropic perpendicular diffusion  coefficient we use in this work. It is the same as the one used in S09, and is taken from Candia \& Roulet (2004).

The diffusion coefficient due to scattering of particles along the mean magnetic field $D_{\parallel}$ is given by
\begin{equation}
D_{\parallel} = c\,L_{\rm max}\,\rho\,\frac{N_{\parallel}}{\sigma^{2}}\,\sqrt{\biggl (\frac{\rho}{\rho_{\parallel}} \biggr )^{2(1 - \gamma)} + 
\biggl (\frac{\rho}{\rho_{\parallel}} \biggr )^{2}} \, ,
\label{eq17}
\end{equation}
where $c$ is the speed of light and the quantities $N_{\parallel}$,
$\gamma$ and $\rho_{\parallel}$ are constants specific for different
kinds of turbulence. The perpendicular diffusion coefficient
($D_{\perp}$) is related to the parallel one ($D_{\parallel}$) by

\begin{equation}
\nonumber
\frac{D_{\perp}}{D_{\parallel}} = \cases{N_{\perp}\,(\sigma^{2})^{a_{\perp}}\, , & $\rho \leq 0.2$ \cr
\noalign{\medskip}
N_{\perp}\,(\sigma^{2})^{a_{\perp}}\,\biggl (\frac{\rho}{0.2} \biggr )^{-2}\, , & $\rho > 0.2$\cr}
\label{eq20}
\end{equation} 

The quantities $N_{\perp}$ and $a_{\perp}$ are constants specific to different kinds of turbulent spectra. In this work, we assume the Kolmogorov turbulence spectrum in our calculations. We use $N_{\parallel}$ = 1.7  , $\gamma$ = 5/3 , $\rho_{\parallel}$ =0.20 , $N_{\perp}$ =0.025  and  $a_{\perp}$= 1.36    ( Table 1,  Candia \& Roulet [2004] ). Eq~(\ref{eq17}) together with Eq~(\ref{eq20}) defines the isotropic perpendicular diffusion  coefficient we use in this work.

\subsection{CME-only model} \label{cme-only}
\begin{figure*}
\centering 
\includegraphics[scale = 0.75]{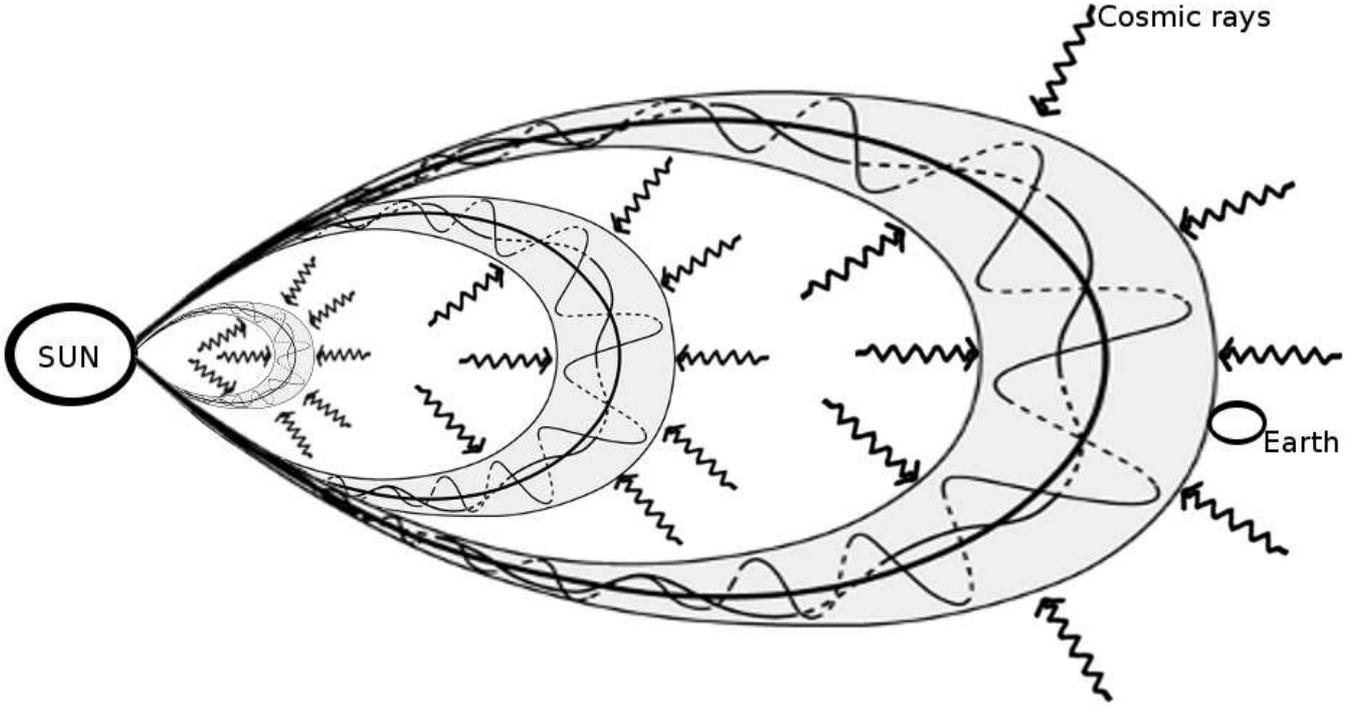}
\caption{A cartoon illustrating a flux rope CME expanding and propagating away from the Sun. High energy galactic cosmic rays diffuse into the CME across its bounding magnetic field.}
\label{crt}
\end{figure*}

The basic features of the CME-only model are similar to that used in S09 and here we only highlight the areas where there are significant differences from S09. There are practically no high-energy galactic cosmic rays inside the CME when it starts out near the Sun. The cosmic rays diffuse into it from the surroundings via perpendicular diffusion across the closed magnetic field lines as it propagates through the heliosphere as shown schematically in Figure~\ref{crt}. Near the Earth, the difference between the (relatively lower) cosmic ray proton density inside the CME and that in the ambient medium appears as the Forbush decrease. We proceed to obtain an estimate of the cosmic ray proton density inside the CME produced by the cumulative effect of diffusion. 

The flux ${\rm F}$ of protons diffusing into the CME at a given time depends on the isotropic perpendicular diffusion coefficient ${\rm D_{\perp}}$ and the density gradient ${\rm \partial N_a/\partial r}$, and can be written as  

\begin{equation}
 \rm {F \,\, {( cm^{-2} s^{-1})}} = \textrm{D}_{\perp} \frac {\partial N_a} {\partial r} \, .
\label{E1}
\end{equation}
 As mentioned earlier, the ${\rm D_{\perp}}$ characterizes diffusion across the (largely closed) magnetic fields bounding the CME and ${\rm N_a}$ is the ambient density of high energy protons. The total number of cosmic ray protons that will have diffused into the CME after a time ${\rm T}$ is related to the diffusing flux by

\begin{equation}
\rm {U}_i = \int_{0}^{T} A(t) F(t) \,dt = \int_{0}^{T}  \textrm{D}_{\perp} A(t) \frac {\partial N_a} {\partial r}\,dt
\label{E2}
\end{equation}

where ${\rm A(t)}$ is the cross-sectional area of the CME at a given time ${\rm t}$. According to our convention, the CME is first observed in the LASCO FOV at ${\rm t=0}$ and it reaches the Earth at ${\rm t=T}$.
The ambient density gradient ${\rm \partial N_{a}/\partial r}$ is approximated by the following expression, which is significantly different from that used in S09:


\begin{equation}
\rm \frac {\partial Na} {\partial r} \simeq \frac{Na}{L}\, ,
\label{E3}
\end{equation}
where $L$ is the gradient lengthscale. Observations of the density gradient lengthscale $L$ exist only  for a few rigidities. We use the observations of Heber et al (2008), who quote a value of $L^{-1}$ = 4.7 \% ${\rm AU}^{-1}$ for 1.2 GV protons. We take this as our reference value. In order to calculate $L$ for other rigidities (in the 14--24 GV range that we use here), we assume that $L \propto R_{L}^{1/3}$. This is broadly consistent with the observation (De Simone et al, 2011) that the density gradient lengthscale is only weakly dependent on rigidity. For a given rigidity, we also need the value of $L$ all the way from the Sun to the Earth. In order to do this, we recognize that $L$ near the CME/magnetic cloud will not be the same as its value in the ambient solar wind. We adopt $ L \propto B_{\rm MC}/B_{\rm a}$, where $B_{\rm MC}$ is the large-scale magnetic field bounding the CME and $B_{\rm a}$ is the (weaker) magnetic field in the ambient medium outside the CME. Furthermore, while $B_{\rm MC}$ varies according to Eq~(\ref{BFL}) below, the ambient field $B_{\rm a}$ of the Parker spiral in the ecliptic plane varies inversely with heliocentric distance.

 We assume that the magnetic flux associated with the CME is ``frozen-in'' with it as it propagates. In other words, the product of the CME magnetic field and the CME cross-sectional area remains constant. This is a fairly well-founded assumption (e.g., Kumar \& Rust 1996; Subramanian \& Vourlidas 2007). One can therefore relate the CME magnetic field ${\rm B_{0}(t)}$ at a given time ${\rm t}$, to the value ${\rm B_{MC}}$ measured in the near-Earth magnetic cloud by
\begin{equation} 
\rm B_0(t) = B_{MC} {\left[ \frac{R(T)}{R(t)} \right]}^2\, ,
\label{BFL}
\end{equation}
where ${\rm R(T)}$ is the radius of the magnetic cloud observed at the Earth and $R(t)$ is its radius at any other time $t$ during its passage from the Sun to the Earth. The CME radius ${\rm R(t)}$ and ${\rm R(T)}$ are related via Eq~(\ref{ER}) below. We emphasize that the magnetic field referred by Eq~(\ref{BFL}) refers to the magnetic field bounding the CME, and not the ambient magnetic field outside it.

We model the CME as an expanding cylindrical flux rope whose length increases with time as it propagates outwards. Its cross-sectional area at time $t$ is
\begin{equation}
\rm A(t) = 2 \pi L(t) R(t)
\label{E4}
\end{equation}
where ${\rm L(t)}$ is the length of the flux-rope cylinder at time ${\rm t}$, and is related to the height ${\rm H(t)}$ of the CME above the solar limb via
\begin{equation}
\rm L(t) = \pi H(t) \, .
\label{E5}
\end{equation}

We note that Eq~(\ref{E5}) differs from the definition used in S09 by a factor of 2. We assume that the CMEs expand in a self-similar manner as they propagate outwards. 3D flux rope fittings to CMEs in the ${\rm \sim 2 - 20}$ ${\rm R_{\odot}}$ field of view using SECCHI/STEREO data validate this assumption ( e.g., Poomvises et al, 2010 ). The self-similar assumption means that the radius of the ${\rm R(t)}$ of the flux rope is related to its heliocentric height ${\rm H(t)}$ by

\begin{equation}
\rm \frac{R(t)}{H(t)} = \frac{R(T)}{H(T)}
\label{ER}
\end{equation}
where ${\rm H(T)}$, the heliocentric height at time ${\rm T}$, is ${\rm  = 1\,AU}$ by definition, and ${\rm R(T)}$ is the measured radius of the magnetic cloud at the Earth.

As mentioned earlier, we consider a 2-stage velocity profile for CME propagation, expressed by Eqs~(\ref{VP1}) and (\ref{VP2}); this is substantially different from the constant speed profile adopted in S09.

Using Eqs~(\ref{E3}), (\ref{E4}) and (\ref{E5}) in (\ref{E2}), we get the following expression for the total number of protons inside the CME when it arrives at the Earth: 

\begin{equation}
\rm \textit{U}_{i} = \int_{0}^{T} 2 \pi L(t) R(t) \textrm{D}_{\perp}  \frac{Na}{\kappa {{R_L(t)} ^{0.33}}} \,dt \, .
\label{E7}
\end{equation}

The cosmic ray density inside the CME when it arrives at the Earth is 
\begin{equation}
\rm \textit{N}_i = \frac {\textit{U}_i}{\pi R(T)^2 L(T)} \, ,
\label{E8}
\end{equation} 
where ${\rm L(T)}$ \ and ${\rm R(T)}$ \ are the length and cross-sectional radius of the CME respectively at time T, when it reaches the Earth. When the CME arrives at the Earth, the relative difference between the cosmic ray density inside the CME and the ambient environment is manifested as the Forbush decrease, whose magnitude ${\rm M}$ can be written as

\begin{equation}
\rm M = \frac { N_a - N_i}{N_a} = \frac { \Delta N}{N_a} \, .
\label{E9}
\end{equation}

We compare the value of the Forbush decrease magnitude ${\rm M}$ predicted by Eq~(\ref{E9}) with observations in \S~\ref{rslt} .

\begin{figure}
   \centering
      \includegraphics[width=0.9\columnwidth]{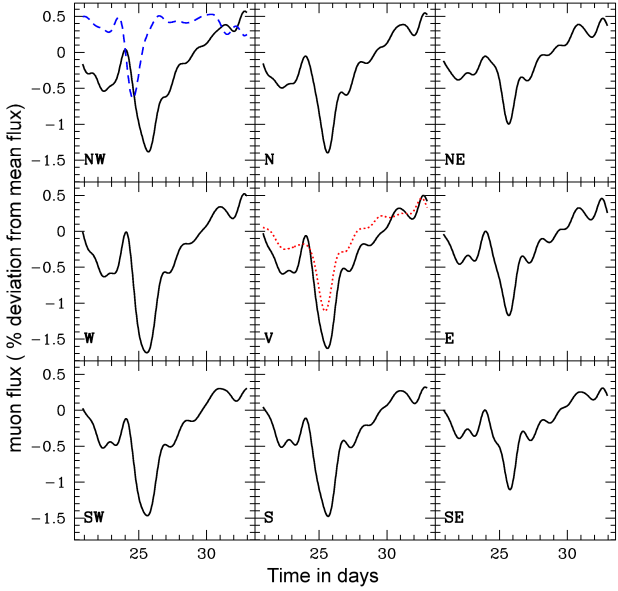}
   \caption{The muon flux along the nine directions is shown for the Forbush decrease on 2001 November 24. The fluxes are shown as percentage deviations from mean values. The solid black lines show the data after applying a low-pass filter (S09). The blue dashed line in the first panel shows the magnetic field observed in-situ by spacecraft. The magnetic field data are inverted (i.e., magnetic field peaks appear as troughs) and are scaled to fit in the panel. The red dotted line in middle panel shows Tibet neutron monitor data scaled down by a factor 3 to fit in the panel.}\label{FD24}%
    \end{figure}

An example of an FD event observed in all 9 bins of GRAPES-3 is shown in Figure~\ref{FD24}. The x-axis is the time in days starting from 1 November 2001 and the y-axis gives the the percentage deviation of the muon flux from the pre-event mean.  The magnitude $M$ of the FD for a given rigidity bin is the difference between the pre-event cosmic ray intensity and the intensity at the minimum of the Forbush decrease.

\subsection{Shock-only model} \label{Sonly}
\begin{figure}
   \centering
      \includegraphics[width = 0.95\columnwidth]{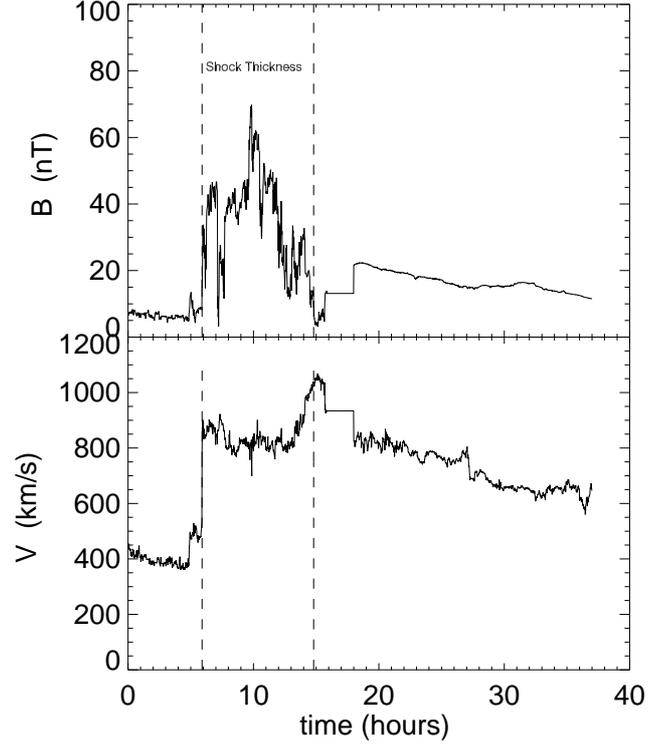}
   \caption{Interplanetary magnetic field and solar wind speed from the day 24 November 2001, The shock sheath thickness is computed by multiplying the time interval inside the dotted lines by the solar wind speed}
              \label{Fshock}%
    \end{figure}

In this approach we assume that the FD is caused exclusively due to the shock, which is modeled as a propagating diffusive barrier.
The expression for the magnitude of the Forbush decrease according to this model is (Wibberenz et al 1998)

\begin{equation}
\rm M \equiv \frac{U_a - U_{shock}}{U_a} = \frac{\Delta U}{U_a} = \frac{V_{sw} L_{shock}}{{D_{\perp}}^a} \left( \frac{{D_{\perp}}^a}{{D_{\perp}}^{shock}} -1 \right)
\label{E14}
\end{equation}

where ${\rm U_a}$  is the ambient cosmic ray density and ${\rm U_{shock}}$ is that inside the shock, ${\rm {D_{\perp}}^a}$ is the ambient isotropic perpendicular diffusion  coefficient and ${\rm {D_{\perp}}^{shock}}$ is that inside the shock, ${\rm V_{sw}}$ is the solar wind velocity and ${\rm L_{shock}}$ is the shock sheath thickness. For each shock event, we examine the magnetic field data from the ACE and WIND spacecrafts and estimate the shock sheath thickness ${\rm L_{shock}}$ to be the spatial extent of the magnetic field enhancement. An example is shown in Figure~\ref{Fshock}.

In computing ${\rm {D_{\perp}}^a}$ and ${\rm {D_{\perp}}^{shock}}$, we need to use different values for the proton rigidity ${\rm \rho}$ for the ambient medium and in the shock sheath; they are related to the proton rigidity $Rg$ by

\begin{eqnarray}
\rm {\rho}^a & = & \frac{Rg}{{B_0}^a L_{shock}} \\
\rm {\rho}^{shock} & = & \frac{Rg}{{B_0}^{shock} L_{shock}} \, ,
\label{E15}
\end{eqnarray}

where $B_{0}^{a}$ is the ambient magnetic field, $B_{0}^{shock}$ is the magnetic field inside the shock sheath.

\section{Results} \label{rslt}
In this section we first describe various parameters needed for the CME-only and the shock-only models that are derived from observations. Using these parameters, we examine whether the notion of cosmic ray diffusion is valid for each model. Using the observationally determined parameters, we then obtain the (CME-only and shock-only) model that best reproduces the observed FD magnitude in each rigidity bin.  
\subsection{Observationally derived parameters}

\begin{figure*}
   \centering
      \includegraphics[width=0.75\paperwidth]{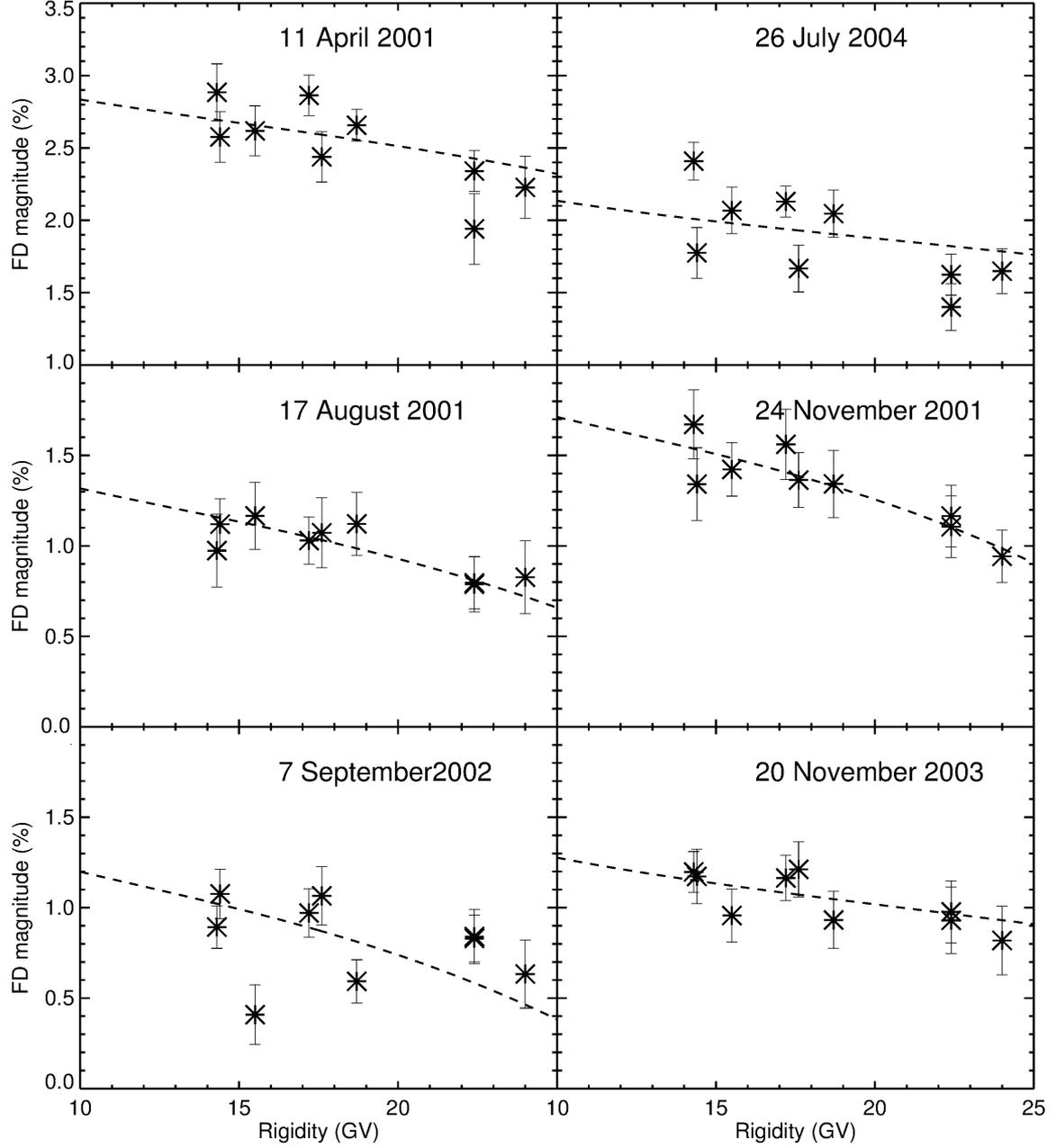}
   \caption{The $\ast$ symbols show the Forbush decrease magnitude observed with GRAPES-3. The dashed line is obtained using the CME-only model}
              \label{Call}%
\end{figure*}

Table~\ref{TCME} contains the observationally determined parameters for each of the CMEs and their corresponding shocks in the final shortlist (Table~\ref{SL3}).

The quantity ``First obs'' denotes the time (in UT) when the CME was first observed in the LASCO FOV, while ${\rm R_{first}}$ is the distance (in units of  ${\rm R_{\odot}}$) at which CME was first observed in LASCO FOV and ${\rm R_{last}}$ is the  distance at which the CME was last observed in LASCO FOV. The quantity ${\rm V_{exp}}$ is the speed of CME at ${\rm R_{last}}$ (in ${\rm km~s^{-1}}$) and  ${\rm a_i}$ is the  acceleration of CME in the LASCO FOV  (in ${\rm m~s^{-2}}$). The quantity ${\rm C_D}$ is the (constant) dimensionless drag coefficient used for the velocity profile (\S~{\ref{VP}}). The quantities MC start and MC end denote the start and end times of the magnetic cloud in UT. The quantity ${\rm V_{sw}^{MC}}$ is solar wind speed at the Earth (in ${\rm km~s^{-1}}$) just ahead of the arrival of the magnetic cloud, and ${\rm R_{MC}}$ is the radius of the magnetic cloud (in ${\rm km }$). The quantity ${\rm B_{MC}}$ is the peak magnetic field inside the magnetic cloud (in ${\rm nT}$). The quantity ${\rm T_{total}}$ is the Sun-Earth travel time (in hours) taken by the CME  to travel from Sun to Earth. 
The quantity Shock arrival denotes the time (in UT) when the shock is detected near the Earth. The quantities ${\rm B^{a}}$ and ${\rm B^{shock}}$ represent the magnetic fields (in nT) in the ambient solar wind and inside shock sheath region respectively (see Fig \ref{Fshock} for an example). The quantity ${\rm V_{sw}^{shock}}$ represents the near-Earth shock speed in ${\rm km~s^{-1}}$. 

Table~\ref{TFD} contains details of the Forbush decreases associated with each of the CMEs in the final shortlist. The magnitude of the FD in a given rigidity bin is the difference between the pre-event intensity of the cosmic rays and the intensity at the minimum of the Forbush decrease. It also contains the onset and the time of minimum, and the magnitude of the decrease in each bin (together with the corresponding cutoff rigidity) for each of the FD events in our final shortlist.

\subsection{Fitting the CME-only and shock-only models to multi-rigidity FD data}

Using the observational parameters listed in Table~\ref{TCME}, we have computed the magnitude of the Forbush decrease using the CME-only (\S~\ref{cme-only}) and shock-only models (\S~\ref{Sonly}). The only free parameter in our model is the ratio of the energy density in the random magnetic fields to that in the large scale magnetic field ${\rm \sigma^2 \equiv \langle {B_{\rm turb}}^2/{B_{0}}^2 \rangle}$. Figure~\ref{Call} shows the best fits of the CME-only model to the multi-rigidity data. The only free parameter in the model is ${\rm \sigma^2 \equiv \langle {B_{\rm turb}}^2/{B_{0}}^2 \rangle}$ , and the best fit is chosen by minimizing the $\chi^2$ with respect to $\sigma^2$. For each FD event, the $\ast$ symbols denote the observed FD magnitude for a given rigidity bin. The dashed line denotes the FD magnitude predicted by the CME-only model.  We define the chi-square statistic as  
\begin{equation}
\rm \chi^2_ = \sum _i \frac {(E_i - D_i)^2} {var_i}
\label{ch2}
\end{equation}
where ${\rm E_i}$ is the value predicted by the theoretical model  ${\rm D_i}$ is the corresponding GRAPES-3 data point and ${\rm var_i}$ is the variance for the corresponding data points. The ${\rm \chi ^2}$ values  obtained after minimizing with respect to $\sigma^2$ are listed in Table~\ref{Chi2}. 

\begin{table}
\caption{{\bf Minimum} ${\rm \chi^2} ${\bf values for the CME-only model fits to GRAPES-3 data}\label{Chi2}}
\centering
\begin{tabular}{l c}
\hline \hline
Event & ${\rm \chi^2}$  \\
\hline
11 April 2001 & 11.1 \\
17 August 2001 & 1.96\\
24 November 2001 & 2.46\\
7 September 2002 & 25.5 \\
20 November 2003 & 3.66 \\
26 July 2004 & 27.5 \\
\hline \hline
\end{tabular}
\end{table}

The entries in the second column $\rm {\sigma_{MC}}$ in Table~\ref{Sig} denote the square roots of the turbulence parameter $\sigma^{2}$ that we have used for the model fits for each event. These values represent the level of turbulence in the sheath region immediately ahead of the CME, through which the cosmic rays must traverse in order to diffuse into the CME. By comparison, the value of $\sigma$ for the quiescent solar wind ranges from 6--15\% (Spangler 2002). Evidently, the CME-only model implies that the sheath region ahead of the CME is only a little more turbulent than the quiescent solar wind, except for the 26 July 2004 event, where the speed of CME at the Earth was much higher than that for the other events.

We have carried out a similar exercise for the shock-only model ({\S~\ref{Sonly}}). For each event, we have used the observationally obtained parameters pertaining to the shock listed in Table~\ref{TCME}. Since this model needs the turbulence levels in both the ambient medium as well as the shock sheath region to be specified, we have assumed that the turbulence level inside the shock sheath region is twice that in the ambient medium. We find that it is not possible to fit the shock-only model to the multi-rigidity data using values for the turbulence parameter that are reasonably close to that in the quiescent solar wind. For each event, last column labeled $\rm {\sigma_{Shock}}$  in Table~\ref{Sig} denotes the turbulence level in the shock sheath region that are required to obtain a reasonable fit to the data. Clearly, these values are an order of magnitude higher than those observed in the quiet solar wind. 

\subsection{${\rm r_{\rm L}/R_{\rm CME}}$ for CME-only model}
\begin{figure*}
   \centering
      \includegraphics[width=0.75\paperwidth]{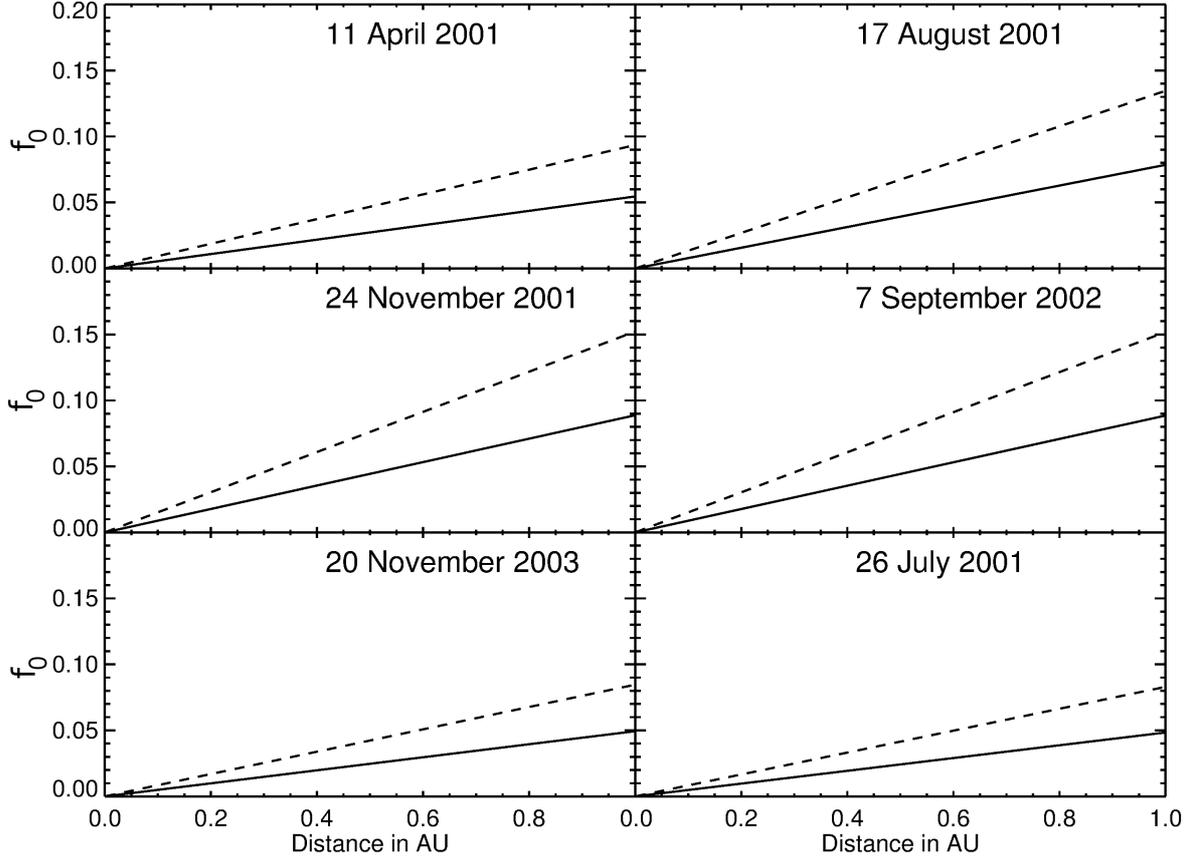}
   \caption{The quantity ${f_{0} \equiv \rm R_{L}/R}$ versus distance as a CME propagates from the Sun to the Earth. The dashed and continuous lines represent 24 GV and 12 GV protons, respectively.}
              \label{Fig1}%
    \end{figure*}

Kubo \& Shimazu (2010) have simulated the process of cosmic ray diffusion into an ideal flux rope CME in the presence of MHD turbulence. They find that, if the quantity ${\rm f_0(t) \equiv R_{L}(t)/R(t)}$ is small, cosmic ray penetration into the flux rope is dominated by diffusion via turbulent irregularities. Other effects such as gradient drift due to the curvature of the magnetic field are unimportant under these conditions. Figure~\ref{Fig1} shows the quantity ${\rm f_0(t) \equiv R_{L}(t)/R(t)}$ for 12 and 24 GV protons, for each of the CMEs in our final shortlist (Table~\ref{SL3}). The Larmor radius ${\rm R_{L}(t)}$ is defined by Eqs~(\ref{rl}) and (\ref{BFL}) and the CME radius ${\rm R(t)}$ is defined in Eq~(\ref{ER}). Clearly, ${\rm f_{0} \ll 1}$ all through the Sun-Earth passage of the CMEs, and this means that the role of MHD turbulence in aiding penetration of cosmic rays into the flux rope structure is expected to be important. 

\section{Summary}
Our main aim in this work is to determine whether Forbush decreases due to cosmic rays of rigidities ranging from  14 to 24 GV are caused primarily by the CME, or by the shock associated with it. We examine this question in the context of multi-rigidity Forbush decrease data from the GRAPES-3 instrument.
We use a carefully selected sample of FD events from GRAPES-3 that are associated with both CMEs and shocks.

We consider two models: the CME-only model (\S~\ref{cme-only}) and the shock-only model (\S~\ref{Sonly}). In the CME-only model, we envisage the CME as an expanding bubble bounded by large-scale magnetic fields. The CME starts out from near the Sun with practically no high energy cosmic rays inside it. As it travel towards the Earth, high energy cosmic rays diffuse into the CME across the large-scale magnetic fields bounding it. The diffusion coefficient is a function of the rigidity of the cosmic ray particles as well as the level of MHD turbulence in the vicinity of the CME (the sheath region). Despite the progressive diffusion of cosmic rays into it, the cosmic ray density inside the CME is still lower than the ambient density when it reaches the Earth. When the CME engulfs the Earth, this density difference causes the Forbush decrease observed by cosmic ray detectors. In the shock-only model, we consider the shock as a propagating diffusive barrier. It acts as an umbrella against cosmic rays, and the cosmic ray density behind the ``umbrella'' is lower than that ahead of it. 

We have obtained a list of Forbush decrease events observed by the GRAPES-3 instrument using the shortlisting criteria described in \S~ \ref{ESC}. For each of these shortlisted events, we have used observationally derived parameters listed in Table~\ref{TCME} for both the models. The only free parameter was the level of MHD turbulence (defined as the square root of the energy density in the turbulent magnetic fluctuations to that in the large-scale magnetic field) in the sheath region. 

\section{Conclusions}
Figure~\ref{Call} shows the results of the CME-only model fits to multi-rigidity data for each of the shortlisted events. For the shock-only model, we use the turbulence level in the shock sheath region as the free parameter. Table~\ref{Sig} summarizes the values of these turbulence levels that we have used for each of the FD events in the final shortlist. These values may be compared with the estimate of 6--15 \% for the turbulence level in the quiescent solar wind (Spangler 2002). We thus find that a good model fit using the CME-only model requires a turbulence level in the sheath region that is typically only a little higher than that in the quiet solar wind, which is generally consistent with observations. On the other hand, a good fit using the shock-only model demands a turbulence level in the shock sheath region that is often an order of magnitude higher than that in the quiet solar wind, which is somewhat unrealistic. The results summarized in Table~\ref{Sig} imply that, for FDs involving protons of rigidities ranging from 14 to 24 GV, the CME-only model is a viable one, while the shock-only model is not. Given the remarkably good fits to multi-rigidity data (Figure~\ref{Call}, the reasonable turbulence levels in the sheath region demanded by the CME-only model (Table~\ref{Sig}) and because the FD minima usually occur well within the magnetic cloud (Table~\ref{TFD}), we conclude that CMEs are the dominant contributors to the FDs observed by the GRAPES-3  experiment.

\begin{table}
\caption{Turbulence levels in the sheath region required by the models}\label{Sig}
\centering
\begin{tabular}{l|cc}
\hline \hline
Event & CME-only model & shock-only model \\
 & ${\rm \sigma _{MC}}$& ${\rm \sigma _{Shock}}$ \\
\hline
11 April 2001 & 9.4 \% & 100 \% \\
17 August 2001 & 13 \%  & 180 \%   \\
24 November 2001 & 28 \%  &   400 \% \\
7 September 2002 & 13\% &100\% \\
20 November 2003 & 6.7 \% &  400 \% \\
26 July 2004 & 46 \%  &  200 \% \\
\hline \hline
\end{tabular}
\end{table}

\begin{acknowledgements}
Arun Babu acknowledges support from a Ph.D studentship at IISER Pune. P. Subramanian acknowledges partial support from the RESPOND program administered by the Indian Space Research Organization. We thank Alejandro Lara for useful discussions and the referee K. Scherer for several critical and helpful suggestions. We thank D. B. Arjunan, A. Jain, the late S. Karthikeyan, K. Manjunath, S. Murugapandian, S. D. Morris, B. Rajesh, B. S. Rao, C. Ravindran, and R. Sureshkumar for their help in the testing, installation, and operating the proportional counters and the associated electronics and during data acquisition. We thank G. P. Francis, I. M. Haroon, V. Jeyakumar, and K.
Ramadass for their help in the fabrication, assembly, and installation of various mechanical components and detectors.
\end{acknowledgements}

\appendix

\section{Tables}
\begin{center}

\begin{table*}
\caption{Derived parameters of Forbush decrease for events,}\label{TFD}
\centering
\begin{tabular}{l|ccccccccc}
\hline \hline 

\multicolumn{10}{l}{\bf 11 April 2001}\\
\multicolumn{10}{l}{$^1$  FD onset times are in UT, 11 April 2001}\\
\multicolumn{10}{l}{$^2$  FD minimum times are in UT, 12 April 2001}\\ \hline
Shock  & Arrival &\multicolumn{8}{l}{  :  11 April 2001, 14:06}\\
Magnetic Cloud  & Start  & \multicolumn{4}{l}{ :  11 April 2001, 23:00} & End &  \multicolumn{3}{l}{ :  12 April 2001, 18:00} \\
\hline
Quantity & NW & N & NE & W & V & E & SW & S & SE  \\
\hline 
FD Magnitude $(\%)$ & 1.02 & 1.12 & 0.91 & 1.38 & 1.40 & 1.05 & 1.36 & 1.29 & 0.93 \\
Cut-off Rigidity $(GV)$ & 15.5  & 18.7 & 24.0 & 14.3 & 17.2 & 22.4 & 14.4 & 17.6 & 22.4 \\
FD Onset $(UT)$ & 12:57 & 13:55 & 16:04 & 11:16 & 12:00 & 12:57 & 08:24 & 10:20 & 12:28 \\
FD minimum $(UT)$ & 19:00 & 19:00 & 19:00 & 17:00 & 18:00 & 18:00 & 14:00 & 15:00 & 16:00 \\ 
\hline \hline
\multicolumn{10}{l}{\bf 17 August 2001}\\
\multicolumn{10}{l}{$^1$  FD onset times are  in UT, 17 August 2001}\\
\multicolumn{10}{l}{$^2$  The $^*$ FD onset times are in UT, 16 August 2001}\\
\multicolumn{10}{l}{$^3$  FD minimum times are in UT, 18 August 2001}\\ \hline
Shock  & Arrival & \multicolumn{8}{l}{ :  17 August 2001, 11:00}\\
Magnetic Cloud  & Start  &  \multicolumn{4}{l}{ :  18 August 2001, 00:00} &End  & \multicolumn{3}{l}{ :  18 August 2001, 21:30} \\
\hline 
Quantity & NW & N & NE & W & V & E & SW & S & SE  \\
\hline
FD Magnitude $(\%)$ & 1.17 & 1.12 & 0.83 & 0.97 & 1.03 & 0.79 & 1.12 & 1.07 & 0.80 \\
Cut-off Rigidity $(GV)$ & 15.5  & 18.7 & 24.0 & 14.3 & 17.2 & 22.4 & 14.4 & 17.6 & 22.4 \\
FD Onset $(UT)$ & 04:19 & 01:55 & 01:12 & 23:17$^*$ & 22:34$^*$ & 22:05$^*$ & 00:00 & 23:31 $^*$& 23:17$^*$ \\
FD minimum $(UT)$ & 4:00 & 3:00 & 2:00 & 6:00 & 5:00 & 5:00 & 23:00 & 22:00 & 20:00 \\
\hline \hline
\multicolumn{10}{l}{\bf 24 November 2001}\\
\multicolumn{10}{l}{$^1$  FD onset times are in UT, 24 November 2001}\\
\multicolumn{10}{l}{$^2$  FD minimum  times are in UT, 25 November 2001}\\ \hline
Shock  & Arrival & \multicolumn{8}{l}{ :  24 November 2001, 06:00}\\
Magnetic Cloud  & Start   &  \multicolumn{4}{l}{ :   24 November 2001, 17:00} & End  & \multicolumn{3}{l}{ :   25 November 2001, 13:00} \\
\hline 
Quantity & NW & N & NE & W & V & E & SW & S & SE  \\
\hline
FD Magnitude $(\%)$ & 1.42 & 1.34 & 0.94 & 1.67 & 1.56 & 1.16 & 1.34 & 1.36 & 1.10 \\
Cut-off Rigidity $(GV)$ & 15.5  & 18.7 & 24.0 & 14.3 & 17.2 & 22.4 & 14.4 & 17.6 & 22.4 \\
FD Onset $(UT)$ & 03:07 & 03:21 & 03:07 & 04:05 & 03:21 & 02:52 & 04:05 & 03:07 & 01:24 \\
FD minimum $(UT)$ & 17:00 & 15:00 & 16:00 & 14:00 & 15:00 & 17:00 & 15:00 & 16:00 & 19:00 \\
\hline \hline
\multicolumn{10}{l}{\bf 7 September 2002}\\
\multicolumn{10}{l}{$^1$ FD onset times are in UT, 7 September 2002}\\
\multicolumn{10}{l}{$^2$ FD minimum times are in UT, 8 September 2002}\\ \hline
Shock  & Arrival & \multicolumn{8}{l}{ :  7 September 2002, 14:20}\\
Magnetic Cloud  & Start    &  \multicolumn{4}{l}{ :   7 September 2002, 17:00} & End  &\multicolumn{3}{l}{ :   8 September 2002, 16:30} \\
\hline 
Quantity & NW & N & NE & W & V & E & SW & S & SE  \\
\hline
FD Magnitude $(\%)$ & 0.41 & 0.59 & 0.63 & 0.89 & 0.97 & 0.83 & 1.08 & 1.07 & 0.84 \\
Cut-off Rigidity $(GV)$ & 15.5  & 18.7 & 24.0 & 14.3 & 17.2 & 22.4 & 14.4 & 17.6 & 22.4 \\
FD Onset $(UT)$ & 17:03 & 17:17 & 16:05 & 14:52 & 14:52 & 15:07 & 15:50 & 16:19 & 17:03 \\
FD minimum $(UT)$ & 15:00 & 14:00 & 14:00 & 13:00 & 13:00 & 14:00 & 15:00 & 16:00 & 16:00 \\
\hline \hline
\multicolumn{10}{l}{\bf 20 November 2003}\\
\multicolumn{10}{l}{$^1$  FD onset times are  in UT, 20 November 2003}\\
\multicolumn{10}{l}{$^2$  The $^*$ FD onset times are in UT, 19 November 2003}\\
\multicolumn{10}{l}{$^3$  The $^{**}$ FD onset times are  in UT, 21 November 2003}\\
\multicolumn{10}{l}{$^4$  FD minimum times are in UT, 24 November 2003}\\  \hline
Shock  & Arrival  & \multicolumn{8}{l}{ :  20 November 2003, 07:30}\\
Magnetic Cloud  & Start    &  \multicolumn{4}{l}{ :   20 November 2003, 10:06} & End  & \multicolumn{3}{l}{ :   21 November 2003, 00:24} \\
\hline 
Quantity & NW & N & NE & W & V & E & SW & S & SE  \\
\hline
FD Magnitude $(\%)$ & 0.95 & 0.93 & 0.81 & 1.19 & 1.16 & 0.97 & 1.17 & 1.20 & 0.93 \\
Cut-off Rigidity $(GV)$ & 15.5  & 18.7 & 24.0 & 14.3 & 17.2 & 22.4 & 14.4 & 17.6 & 22.4 \\
FD Onset $(UT)$ & 21:22 $^*$ & 08:10$^{**}$ & 02:53$^{**}$ & 01:55 & 10:48 & 20:38 & 06:58 & 10:19 & 15:07 \\
FD minimum $(UT)$ & 5:00 &  4:00 &  2:00 &  4:00 &  4:00 &  4:00 &  3:00 &  3:00 & 2:00 \\
\hline \hline
\multicolumn{10}{l}{\bf 26 July 2004}\\
\multicolumn{10}{l}{$^1$  FD onset times are  in UT, 26 July 2004}\\
\multicolumn{10}{l}{$^2$  FD minimum times are  in UT, 27 July 2004}\\ \hline
Shock  & Arrival  & \multicolumn{8}{l}{ :  26 July 2004, 22:20}\\
Magnetic Cloud  & Start   &  \multicolumn{4}{l}{ :   27 July 2004, 02:00} & End & \multicolumn{3}{l}{ :   27 July 2004, 24:00} \\
\hline 
Quantity & NW & N & NE & W & V & E & SW & S & SE  \\
\hline
FD Magnitude $(\%)$ & 2.06 & 2.04 & 1.67 & 2.40 & 2.12 & 1.62. & 1.77 & 1.67 & 1.40 \\
Cut-off Rigidity $(GV)$ & 15.5  & 18.7 & 24.0 & 14.3 & 17.2 & 22.4 & 14.4 & 17.6 & 22.4 \\
FD Onset $(UT)$ & 14:24 & 15:22 & 17:17 & 14:10 & 15:36 & 18:00 & 16:05 & 18:29 & 20:38 \\
FD minimum $(UT)$ & 10:00 & 11:00 & 13:00 & 10:00 & 11:00 & 13:00 & 12:00 & 14:00 & 15:00 \\
\hline \hline

\end{tabular}
\tablefoot{
For each event, first row: magnitude of Forbush decrease (FD magnitude), second row: cut-off rigidity, third row: Forbush decrease onset time, fourth row: FD minimum time. 
The units in which each quantity is expressed is given in parentheses in the first column.}
\end{table*}
\end{center}

\begin{table*}
\caption{Observed  parameters of  CME $\&$ Shock for different events}\label{TCME}
\centering
\begin{tabular}{l|cccccc}
\hline \hline
 Event & 11 Apr 2001 & 17 Aug 2001 & 24 Nov 2001 & 7 Sep 2002 & 20 Nov 2003 & 26 Jul 2004 \\ \hline \hline 
\multicolumn{7}{l} { CME details }\\ \hline
 \tablefootmark{a}First obs.   ( UT) & $10/04/01$, 05:30 & $15/08/01$, 23:54 & $22/11/01$, 22:48 & $5/09/02$, 16:54 & $18/11/03$, 8:50 & $25/07/04$, 14:54 \\ \hline
\tablefootmark{b}${\rm R_{first}}$ $\, (R_{\odot})$ & 2.84  & 3.38 & 4.77  & 4.12  & 6.3  & 4.22 \\ \hline
\tablefootmark{c}${\rm R_{last}}$  $\, (R_{\odot})$ & 18.1  & 25.9 & 25.9  & 17.0 & 27.5 & 21.9\\ \hline
\tablefootmark{d}${\rm V_{exp}}$  $(km\, s^{-1})$& 2880 & 1410 & 1370 & 1860 & 1650 & 1370 \\ \hline
\tablefootmark{e}${\rm a_i}$ $(m\, s^{-2})$ & 211 & -31.7 & -12.8 &  43.0 & -3.29 & 7.0 \\  \hline
\tablefootmark{f}${\rm C_D}$ & 0.325 & 0.163 & 0.09 & 0.312 & 0.333 & 0.016 \\ \hline
\tablefootmark{g}MC start   (UT) & $11/04/01$  , 23:00 & $18/08/01$, 00:00 & $24/11/01$, 17:00 & $7/09/02$, 17:00 & $20/11/03$, 10:06 & $27/07/04$, 2:00 \\ \hline
\tablefootmark{h}MC end    (UT)& $12/04/01$, 18:00 &  $18/08/01$, 21:30 & $25/11/01$, 13:00 & $8/09/02$, 16:30 & $21/11/03$, 00:24 & $27/07/04$, 24:00  \\ \hline
\tablefootmark{i}${\rm V_{sw}^{MC}}$ $(km\, s^{-1})$ & 725 & 600  & 730 & 544 & 750 & 900 \\ \hline
\tablefootmark{j}${\rm R_{MC}}$  $(km\,) $& $2.48 \times 10^7 $  & $2.32 \times 10^7 $ & $2.63 \times 10^7 $ & $2.30 \times 10^7 $ & $1.89 \times 10^7 $ & $3.56 \times 10^7 $\\ \hline
\tablefootmark{k}${\rm B_{MC}}$   $(nT)$ & 34.5  & 25.6  & 20  & 22.9 & 50  & 25.3  \\ \hline
\tablefootmark{l}${\rm T_{total}}$ $(hours)$ & 42.1  & 44.9 & 41.1 & 48.9  & 42.1 & 35.7 \\ \hline \hline
\multicolumn{7}{l} {Shock }\\ \hline
Shock arrival (UT)& $11/04/01$, 14:06 & $17/08/01$, 11:00 & $24/11/01$, 6:00 & $7/09/02$, 14:20 & $20/11/03$, 7:30 & $26/7/04$, 22:20 \\ \hline
\tablefootmark{m}${\rm B^{a}}$ $(nT)$ & 4.5 & 5 & 5 & 5.8 & 7 & 5 \\ \hline
\tablefootmark{n}${\rm B^{shock}}$ $(nT)$& 32.5 & 33 & 41.5 & 23 & & 26.1\\ \hline
\tablefootmark{o}${\rm V_{sw}^{shock}}$  $(km\, s^{-1})$& 670 & 501 & 948 & 550 & & 893\\ \hline \hline
\end{tabular}
\tablefoot{
\tablefoottext{a}{First obs. : time in UT when the CME was first observed}\\
\tablefoottext{b}{${\rm R_{first}}$  : distance in  $\, (R_{\odot})$, where CME was first observed.}\\
\tablefoottext{c}{${\rm R_{last}}$ : distance in   $\, (R_{\odot})$, where the CME was last observed in the LASCO FOV.}\\
\tablefoottext{d}{${\rm V_{exp}}$  : speed of CME in $(km\, s^{-1})$ at ${\rm R_{last}}$ }\\
 \tablefoottext{e}{${\rm a_i}$  : acceleration of CME in LASCO FOV in units of $(m\, s^{-2})$ }\\
\tablefoottext{f}{${\rm C_D}$ : the constant dimensionless drag coefficient used for the velocity profile (\S~2.2).}\\
\tablefoottext{g}{MC start :  Magnetic cloud start time in UT. }\\
\tablefoottext{h}{MC end : Magnetic cloud end time in UT.}\\
\tablefoottext{i}{${\rm V_{sw}^{MC}}$ : solar wind speed in $(km\, s^{-1})$ at Earth during the arrival of the magnetic cloud}\\
\tablefoottext{j}{${\rm R_{MC}}$ : radius of magnetic cloud in  $(km\,) $}\\
 \tablefoottext{k}{${\rm B_{MC}}$ : magnetic field inside the magnetic cloud  $(nT)$}\\
\tablefoottext{l}{${\rm T_{total}}$ : The Sun-Earth travel time {\em in hours} for the CME.}\\
\tablefoottext{m}{${\rm B^{a}}$ : magnetic field in the ambient solar wind (in $nT$ )}\\
\tablefoottext{n}{${\rm B^{shock}}$ : magnetic field inside the shock sheath region (in $nT$)}\\
\tablefoottext{o}{${\rm V_{sw}^{shock}}$ : near-Earth shock speed $(km\, s^{-1})$}\\ }
\end{table*}

\end{document}